\documentclass[a4paper,11pt]{article}
\usepackage{pos}

\title{MeV neutrino dark matter in the SLIM model}
\ShortTitle{MeV dark matter in the SLIM model}

\author*[a]{Juri Fiaschi}
\author[a]{Michael Klasen}
\author[b]{Miguel Vargas}
\author[b]{Christian Weinheimer}
\author[a]{Sybrand Zeinstra}

\affiliation[a]{Institut für Theoretische Physik, Westfälische Wilhelms-Universität Münster, Wilhelm-Klemm-Straße 9, 48149 Münster, Germany}
\affiliation[b]{Institut für Kernphysik, Westfälische Wilhelms-Universität  Münster, Wilhelm-Klemm-Straße 9, 48149 Münster, Germany}

\emailAdd{fiaschi@uni-muenster.de}
\emailAdd{michael.klasen@uni-muenster.de}
\emailAdd{weinheimer@uni-muenster.de}
\emailAdd{m\_varg03@uni-muenster.de}
\emailAdd{swzeinstra@uni-muenster.de}

\abstract{
We explore the parameter space of a variant of the SLIM model, which extends the SM with a singlet and a doublet of complex scalars and two generations of right-handed neutrinos, the lightest of which has a mass in the MeV to GeV region and plays the role of Dark Matter candidate.
We impose the current collider and astrophysical constrains, as well as bounds from Lepton Flavour Violating experiments.
We also consider the discovery potential in the XENON experiment exploiting the electron recoil as a possible direct detection signal.
Despite the DM in this model being leptophilic, the predicted cross sections are too low due to the heavy charged mediator.
}

\FullConference{%
  40th International Conference on High Energy physics - ICHEP2020\\
  July 28 - August 6, 2020\\
  Prague, Czech Republic (virtual meeting)
}

\begin{document}

\begin{flushright}
MS-TP-20-39\\
\end{flushright}

\maketitle

\section{Introduction}
\vspace{-1em}
\noindent

While its specific nature is still unclear, the presence of Dark Matter (DM) in our universe has been corroborated by several experimental evidences, and its relic density has been measured very precisely by the Planck~\cite{Aghanim:2018eyx}.
While cold DM with mass in the GeV to TeV range has been used to explain the large structure of the universe, warm DM such as sterile neutrinos can better account for other cosmological observations, such as the missing satellite galaxies, the cusp-core problem of inner DM density profiles and the too-big-to fail problem.
Recent measurements in the neutrino sector, such as the solar and atmospheric neutrino mass difference~\cite{Tanabashi:2018oca}, also call for physics Beyond the Standard Model (BSM).

A common theoretically well justified extension of the SM consists in the inclusion of right-handed neutrinos, which can generate small neutrino masses through various realisations of the see-saw mechanism.
When this mechanism is realised at one loop, the neutrino and DM sectors become connected, leading to interesting phenomenological signatures.
BSM constructions which explore such possibility are called Minimal Models, and several examples can be found in the literature.
Recently these models have also been classified according to their neutrino masses generating loop topologies~\cite{Restrepo:2013aga}.

Here we consider a variant of the SLIM (Scalar as Light as MeV) model~\cite{Boehm:2006mi,Farzan:2009ji} with a light sterile right-handed neutrino playing the role of DM candidate with mass in the MeV region, and were the correct neutrino masses are generated radiatively~\cite{Fiaschi:2019evv}.

\section{The SLIM model with MeV neutrino DM}
\vspace{-1em}
In this version of the SLIM model, the particle content of the SM is augmented by an extra complex scalar singlet field ($\rho$), a complex scalar doublet ($\eta$) and by two generations of Majorana right-handed neutrinos ($N_i$, $i$ = 1,2).
The new fields are stabilised by an extra global $U(1)$ symmetry, which gets softly broken to $Z_2$.

The Lagrangian with the Higgs ($\Phi$) potential and the new terms is
\vspace{-0.5em}
\begin{eqnarray}
 \mathcal{L} &=& - m_1^2 \Phi^\dagger \Phi - m_2^2 \eta^\dagger \eta - m_3^2 \rho^*\rho - \frac{1}{2} m_4^2 \left(\rho^2+ (\rho^*)^2 \right) - \mu (\eta^\dagger \Phi \rho + h.c.) - \frac{1}{2}m_{N_i} \overline{N^c_i}N_i \nonumber\\
    && - \frac{1}{2}\lambda_1 (\Phi^\dagger\Phi)^2 - \frac{1}{2}\lambda_2 (\eta^\dagger\eta)^2  - \frac{1}{2}\lambda_3 (\rho^*\rho)^2 - \lambda_4 (\eta^\dagger\eta)(\Phi^\dagger\Phi)- \lambda_5 (\eta^\dagger\Phi)(\Phi^\dagger\eta) \nonumber\\
    && - \lambda_6 (\rho^*\rho)(\Phi^\dagger\Phi)- \lambda_7 (\rho^*\rho)(\eta^\dagger\eta)  - \left(\lambda_8 \right)_{ij} (\overline{N^c_i} \eta^\dagger L_j + h.c.).
\end{eqnarray}
\noindent
The parameter $m_4$ is assumed to be small as it is responsible for the soft breaking of $U(1)$ into $Z_2$.
The new scalars acquire mass after electroweak symmetry breaking.
The charged components will have mass
\begin{equation}
 m_{\eta^\pm}^2 = m_2^2 + \frac{1}{2}\lambda_4 v^2
\end{equation}
\noindent
while the neutral components of the singlet and doublet mix and their mass matrix reads
\begin{equation}
 M_{R,I}^2 = \begin{pmatrix}
    m_2^2 + (\lambda_4 + \lambda_5) \frac{v^2}{2} & \mu \frac{v}{\sqrt{2}} \\
    \mu \frac{v}{\sqrt{2}} & m_3^2 + \lambda_6 \frac{v^2}{2} \pm m_4^2
    \end{pmatrix} =:
 \begin{pmatrix}A & \!\!\!\!B \\ B & \ C_{R,I} \end{pmatrix}
\end{equation}
\noindent
with the positive and negative signs associated to the real and imaginary components respectively.
The mass splitting between the two is thus small, and the masses of their eigenstates is

\begin{equation}
 m_{R,I}^2 = \frac{1}{2}\left(A+C_{R,I} \pm \sqrt{(A-C_{R,I})^2 +4B^2}\right).
\end{equation}
\noindent
We parametrise their mass difference by introducing the parameter $\epsilon$
\vspace{-0.5em}
\begin{equation}
 AC_{R,I}-B^2=:\epsilon(A+C_{R,I})
\end{equation}
\noindent
By tuning this small parameter we obtain two MeV eigenvalues, while the other two remain as heavy as the charged scalars.

\section{Experimental constrains}
\vspace{-1em}
The large parameter space of the model is strongly narrowed by experimental constrains.
The masses of the new scalars, thus the parameters $m_{2,3}$, and the couplings $\lambda_{4,5,6}$ are constrained by ATLAS~\cite{Aaboud:2019rtt} and CMS~\cite{Sirunyan:2018owy} limits on the Higgs branching ratio into invisible decays.
The masses of the charged scalars are excluded below 98.5 GeV by LEP measurements~\cite{Abbiendi:2003yd} and their couplings to the Higgs are restricted below a certain threshold by measurements on the Higgs to two photon branching ratio~\cite{Sirunyan:2018koj}.
The choice $\lambda_{4,5} \simeq$ 0.1 satisfies the above constrains, and at the same time provides with a relatively large electron recoil cross section.

To provide for a valid solution for the cosmological problems cited above, the masses of the singlet and of the doublet neutral eigenstates shall be close.
For a better control on the mass splitting, we fix $\lambda_6$ = 2.3, while we vary the parameter $\epsilon$ between 10$^{-5}$ and 60 GeV$^2$.
Moreover, in order not to erase primordial DM fluctuation, for MeV DM and weak DM self-interactions, we need a small mass difference between the DM particle and the lightest scalar.
For this purpose we vary the ratio of the right handed neutrino masses over the lighter scalars ones between 0.1 and 0.98.
Finally, the couplings $\lambda_{2,3,7}$ have a small impact on the phenomenology, and they are fixed $\simeq$ 0.1 - 0.2.

Because of the residual $Z_2$ symmetry, no tree-level see-saw mechanism is allowed.
The neutrino masses are generated at one-loop through the interaction of the new neutral scalar fields and the right-handed neutrinos.
In particular, similarly to the scotogenic model, the neutrino masses arise from the small mass splitting between the real and imaginary scalars, and they are directly proportional to the coupling $\lambda_8$.

We adopt the very convenient Casas-Ibarra parametrisation~\cite{Casas:2001sr}, which allows to take as input the experimentally measured values of the neutrino mass differences and mixing angles of the PMNS matrix, together with the dark particles masses, and it returns the associated value of $\lambda_8$.
With this method the parameters of the neutrino sector are automatically satisfied for any point in the parameter space of the model.

\section{Numerical Results}
\begin{figure}
 \centering
 \includegraphics[width=0.46\textwidth]{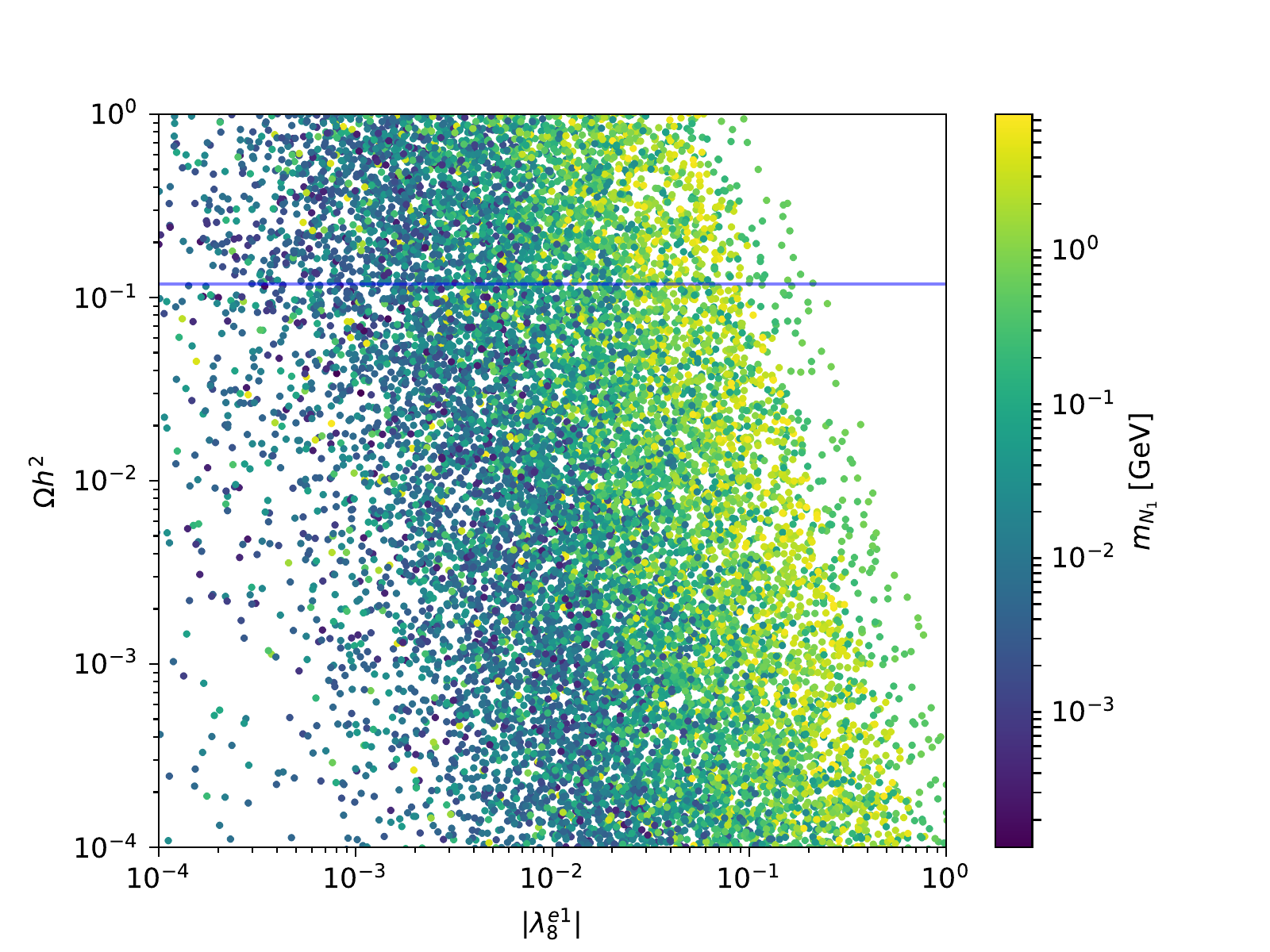}{(a)}
 \includegraphics[width=0.46\textwidth]{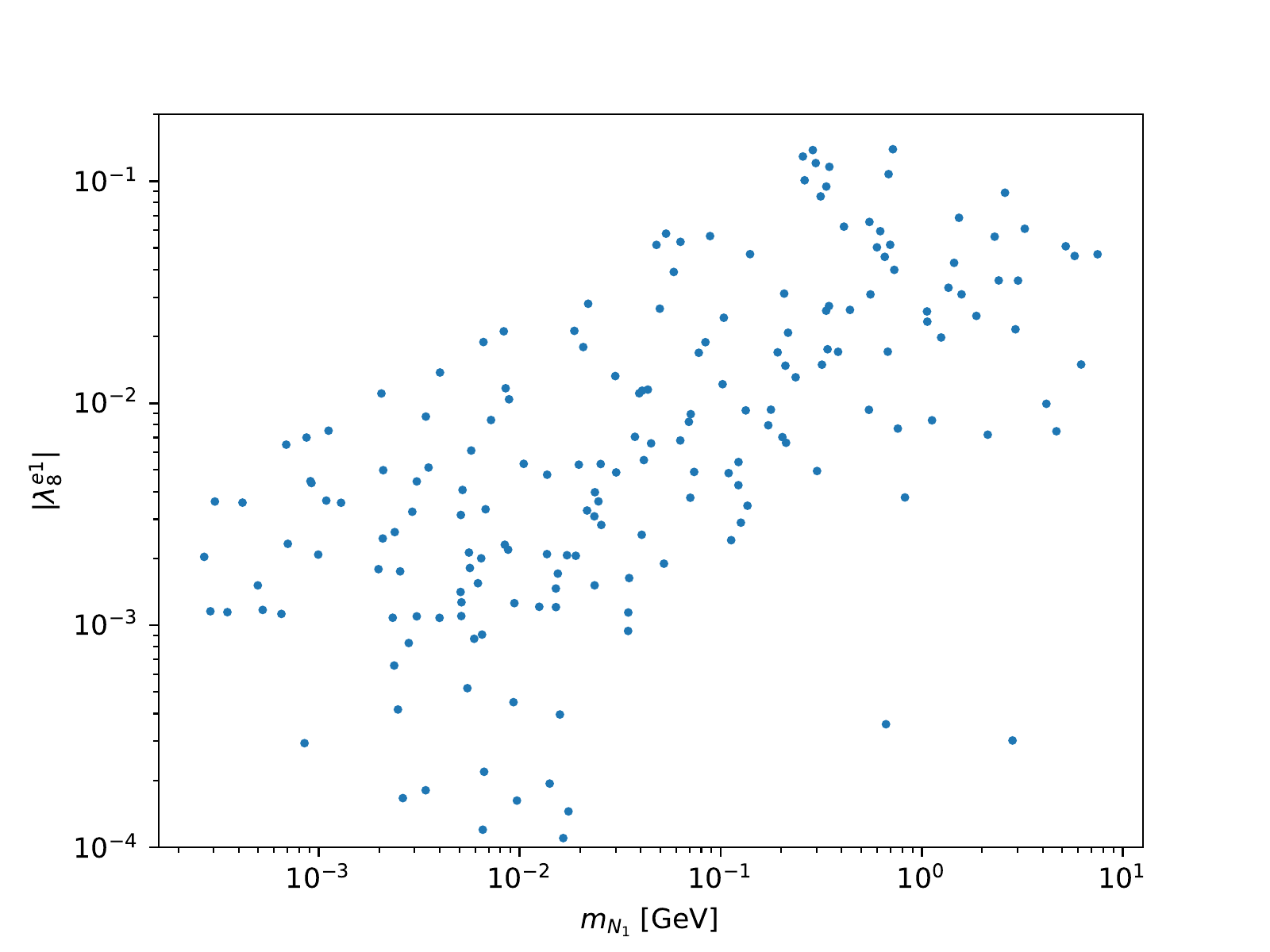}{(b)}
 \caption{(a) The relic density $\Omega h^2$ in the plane right-handed neutrino DM mass versus the absolute value of its coupling to electrons and electron neutrinos. The relic density measured by Planck~\cite{Aghanim:2018eyx}
 is shown as a blue horizontal band, and the DM mass is given on a colour scale.
 (b) The projection of the points satisfying all constrains, including correct relic density, in the plane DM-electron coupling versus DM mass.}
\label{fig:Relic_density}
\end{figure}

\vspace{-1em}
We obtained the following numerical results scanning over appropriate regions of the model's parameter space.
Fig.~\ref{fig:Relic_density} on the left shows the right-handed neutrino DM relic density that is obtained for the parameter space points with certain electrons to electron neutrinos coupling $\lambda_8^{e1}$. The color scheme shows the respective value of the DM mass.
The measured relic density is represented by the blue horizontal line, and it can be generated by various appropriate combinations of couplings and DM masses.
In particular, as the coupling rises, the DM mass increases from the MeV to the GeV region.
In the figure on the right is shown the projection of the model's parameter space points leading to the correct relic density (and satisfying the other experimental constrains discussed above) in the plane of DM-electron coupling and DM mass.
Again, larger couplings, thus more efficient DM annihilation processes, are in order to reduce the DM abundance for heavier DM.

\begin{figure}
 \centering
 \includegraphics[width=0.46\textwidth]{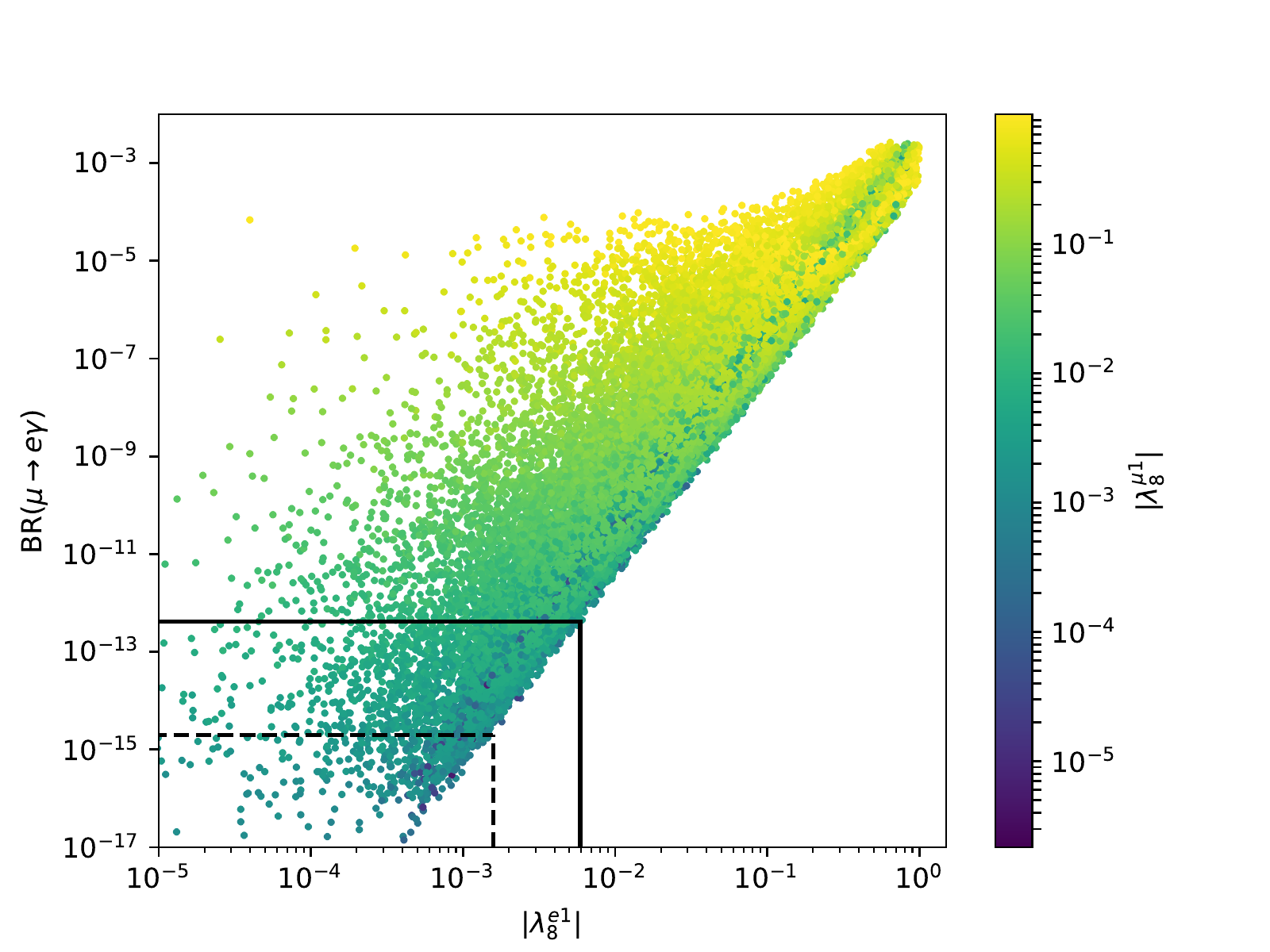}{(a)}
 \includegraphics[width=0.46\textwidth]{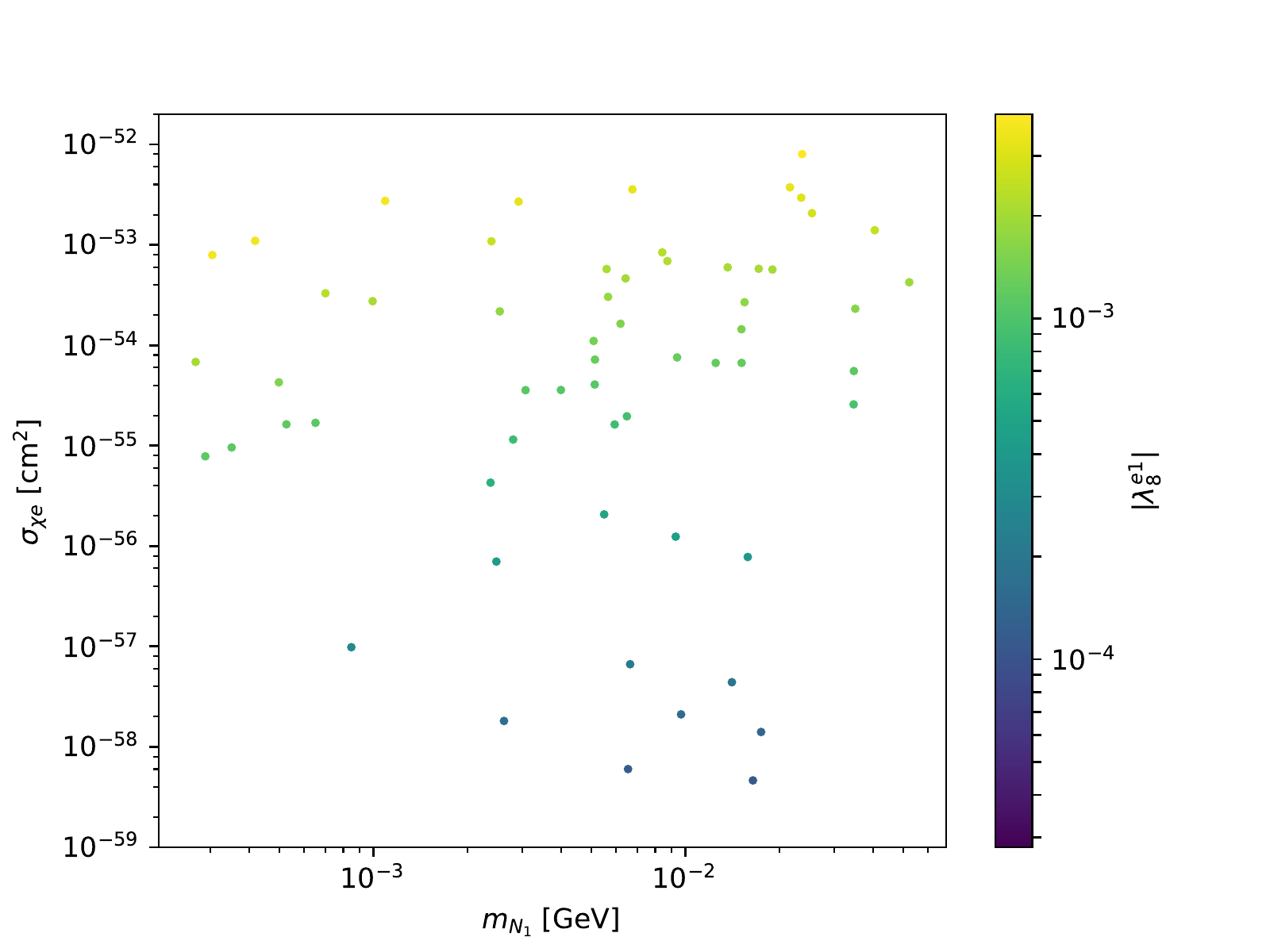}{(b)}
 \caption{(a) Branching ratio of the lepton flavour violating process
 $\mu \rightarrow e \gamma$ versus the DM-electron coupling $\lambda_8^{e1}$.
 The DM-muon coupling $\lambda^{\mu 1}_8$ is shown on a colour scale.
 Solid and dashed black lines represent the current \cite{TheMEG:2016wtm} and expected future~\cite{Renga:2018fpd} limits imposed by the MEG experiment.
 (b) Scattering cross section of right-handed neutrino DM off electrons versus the DM mass. The DM-electron coupling $\lambda_8^{e1}$ is shown on a colour scale.}
 \label{fig:Lepton_flavour_violation_Electron_recoil_XS}
\end{figure}

The right-handed neutrinos are thus connected to charged leptons through sizeable coupling $\lambda_8$, and in this context Lepton Flavour Violating (LFV) processes occur at one-loop.
The experimentally most sensitive LFV process being the flavour-changing neutral current $\mu \rightarrow e\gamma$, sets strong bounds on the $\lambda_8^{ei}$ and $\lambda_8^{\mu i}$ couplings, which in turn were connected to the parameters of the neutrino section through the Casas-Ibarra parametrisation.
The BR for this process is shown in Fig.~\ref{fig:Lepton_flavour_violation_Electron_recoil_XS}(a) as function of $\lambda_8^{e1}$ in the $x$-axis and of $\lambda_8^{\mu 1}$ in the colour scale.
The black solid and dashed lines represent the current~\cite{TheMEG:2016wtm} and expected future~\cite{Renga:2018fpd} limits imposed by the MEG experiment, and as visible they already exclude large part of the model's parameter space, forcing $\lambda_8^{e1}$ below 6 $\times$ 10$^{-3}$ and $\lambda_8^{\mu 1}$ below 10$^{-2}$.
Future constrains will lower these threshold further by roughly a factor four.
The experimental limits on other LFV processes carry similar but somewhat slightly weaker constrains on the same couplings.

\section{Sensitivity on electron recoil}
\vspace{-1em}
The relatively large couplings between the DM right-handed neutrinos and charged leptons make the SLIM model an ideal candidate to explore the sensitivity of direct detection experiment in their novel electron-recoil analysis.

A generalisation of the usual scattering over nuclei analysis has been proposed by the XENON collaboration~\cite{Aprile:2016wwo} to lower their detector energy threshold down to few KeV by abandoning the scintillation light requirement S1 and using only the charge signal S2.
The disadvantage of this method is that fiducialisation is limited, since without S1 the event depth $z$ cannot be accurately estimated, yielding a potentially increased background.

In an optimal configuration, the XENON100 experiment could became sensitive to DM masses down to 600 MeV with a lowest cross section of 6$\times$10$^{-35}$ cm$^2$ for a DM mass of 2 GeV and axial-vector couplings.
Very recently XENON1T published a S2-only analysis~\cite{Aprile:2019xxb} with an electron energy threshold of 0.4 keV, which would be sensitive to light DM scattering off electrons down to masses of 100 to 20 MeV.

In Fig.~\ref{fig:Lepton_flavour_violation_Electron_recoil_XS}(b), we show the DM-electron scattering cross section in the SLIM model for the parameter space points satisfying all the collider, LFV and astrophysical constrains, as function of the DM mass and with colour scheme reflecting the size of the $\lambda_8^{e1}$ coupling.
Even for the largest possible coupling ($\lambda_8^{e1} \geq$ 0.1), the cross section is at most of the order of 10$^{-46}$ cm$^2$, while when further imposing the LFV constrains it is bounded below 10$^{-52}$ cm$^2$ as visible in the plot.
The strong suppression is due to the large mass of the mediator $\eta^{\pm}$, which enters with the fourth power in the calculation of the electron recoil cross section.

\section{Conclusions}
\vspace{-1em}
We have analysed a variant of the SLIM model, predicting MeV right-handed neutrino DM where the particles in the dark sector are responsible for the small neutrino masses which are generated at one-loop.
We have shown that collider, astrophysical and LFV constrains drastically restrict the allowed parameter space of the model, where however various viable points can be found.
The leptophilic nature of the model allows for potential discovery in the XENON1T direct detection experiment exploiting the recently proposed analysis of electron recoil signals by the collaboration.
We have verified that once all experimental constrains are taken into account, the size of the DM-electron cross sections of the viable points are several orders of magnitude below the sensitivity of the experiment.

\section*{Acknowledgements}
\vspace{-1em}
\noindent
We thank J.\ Alvey for useful discussions.
This work has been supported by the BMBF under contract 05H18PMCC1 and the
DFG through the Research Training Group 2149 ``Strong and weak interactions
-- from hadrons to dark matter''.

\end{document}